\documentstyle[prd,aps,epsfig]{revtex}
\bibstyle{unsrt}
\begin{document}
\preprint{LAUR 99-5263}
\twocolumn[\hsize\textwidth\columnwidth\hsize\csname 
@twocolumnfalse\endcsname
\draft

\title{A Second-Order Stochastic Leap-Frog Algorithm for
Multiplicative Noise Brownian Motion} 
\author{Ji Qiang$^{1,\star}$ and Salman Habib$^{2,\dagger}$}
\address{$^1$LANSCE-1, MS H817, Los Alamos National Laboratory, Los
Alamos, NM 87545} 
\address{$^2$T-8, Theoretical Division, MS B285, Los Alamos National
Laboratory, Los Alamos, NM 87545}
\date{\today}
\maketitle

\begin{abstract}
A stochastic leap-frog algorithm for the numerical integration of
Brownian motion stochastic differential equations with multiplicative
noise is proposed and tested. The algorithm has a second-order
convergence of moments in a finite time interval and requires the
sampling of only one uniformly distributed random variable per time
step. The noise may be white or colored.  We apply the algorithm to a
study of the approach towards equilibrium of an oscillator coupled
nonlinearly to a heat bath and investigate the effect of the
multiplicative noise (arising from the nonlinear coupling) on the
relaxation time. This allows us to test the regime of validity of the
energy-envelope approximation method.
\end{abstract}

\pacs{PACS Numbers : 05.10.-a, 05.40.-a, 02.60.Cb, 02.50.Ey
\hfill   LAUR 99-5263} 

\vskip2pc]

\section{Introduction}

Stochastic differential equations with multiplicative noise have not
only found many applications in physics but also have interesting
mathematical properties. Consequently they have attracted substantial
attention over the
years~\cite{zwanzig,lindenberg,careta,habib1,habib2,efremov,becker,leung,bao,mangioni,genovese}.
The key point lies in the fundamental difference between additive and
multiplicative noises: Additive noise does not couple directly to the
system variables and disappears from the noise-averaged form of the
dynamical equations. However, in the case of multiplicative noise, the
system variables do couple directly to the noise (alternatively, we
may say that the noise amplitude depends on the system
variables). This fact can lead to dramatic changes of system behavior
that cannot occur in the presence of additive noise alone. Two classic
illustrations are the Kubo oscillator~\cite{kosc} and the existence of
long-time tails in transport theory \cite{rwzltt}. In this paper we
will investigate another example, that of an oscillator nonlinearly
coupled to a heat bath, in which the effects of multiplicative noise
can significantly alter the qualitative nature, as well as the
rate~\cite{lindenberg}, of the equilibration process (relative to that
of an oscillator subjected only to additive noise).

The dynamical behavior of systems subjected to noise can be studied in
two different ways: we may either solve stochastic differential
equations and average over realizations to obtain statistical
information, or we may directly solve the Fokker-Planck equation which
describes the evolution of the corresponding probability distribution
function. Both approaches have their share of advantages and
disadvantages. Fokker-Planck equations are partial differential
equations and their mathematical properties are still not fully
understood. Moreover, they are very expensive to solve numerically
even for dynamical systems possessing only a very modest number of
degrees of freedom. Truncation schemes or closures (such as cumulant
truncations) have had some success in extracting the behavior of
low-order moments, but the systematics of these approximations remains
to be elucidated. Compared to the Fokker-Planck equation, stochastic
differential equations are not difficult to solve, and with the advent
of modern supercomputers, it is possible to run very large numbers of
realizations in order to compute low-order moments accurately. (We may
mention that in applications to field theories it is essentially
impossible to solve the corresponding Fokker-Planck equation since the
probability distribution is now a functional.)  However, the
extraction of the probability distribution function itself is very
difficult due to the sampling noise inherent in a particle
representation of a smooth distribution.

Numerical algorithms to solve stochastic differential equations have
been discussed extensively in the
literature~\cite{greiner,mannella1,mannella2,honeycutt,kloeden,mannella3}.
The simplest, fastest, and still widely-used, is Euler's method which
yields first-order convergence of moments for a finite time interval.
Depending on the control over statistical errors arising from the
necessarily finite number of realizations, in the extraction of
statistical information it may or may not pay to use a higher order
algorithm especially if it is computationally expensive. Because of
this fact, it is rare to find high-order schemes being put to
practical use for the solution of stochastic differential equations,
and second-order convergence is usually considered a good compromise
between efficiency and accuracy.  A popular algorithm with
second-order convergence of moments for additive noise but with only
first-order convergence of moments for multiplicative noise is Heun's
algorithm (also called stochastic RK2 by some
authors)~\cite{greiner,honeycutt,habib3}.  A stochastic leap-frog
algorithm which has the same order convergence of moments as Heun's
method was suggested in Ref.~\cite{seebelberg} to study particle
motion in a stochastic potential without damping.  Several other
algorithms for particle motion in a quasi-conservative stochastic
system were proposed in Ref. \cite{mannella2} and in the book by Allen
and Tildesley~\cite{allen}.  At every time step, these methods all
require sampling two Gaussian random variables which adds to the
computational cost. A modified algorithm suggested in
Ref. \cite{mannella3} requires only one Gaussian random variable but
applies only to white Gaussian noise. In the following sections, we
present a new stochastic leap-frog algorithm for multiplicative
Gaussian white noise and Ornstein-Uhlenbeck colored noise which not
only has second-order convergence of moments but also requires the
sampling of only one random uniform variable per time step.

The organization of this paper is as follows: General numerical
integration of a system of stochastic differential equations with
Gaussian white noise is discussed in Section~II. The stochastic
leap-frog algorithms for Brownian motion with Gaussian white noise and
colored Ornstein-Uhlenbeck noise are given in Section~III. Numerical
tests of these algorithms using a one-dimensional harmonic oscillator
are presented in Section~IV. A physical application of the algorithm
to the multiplicative-noise Brownian oscillator is given in
Section~V. Section~VI contains the final conclusions and and a short
discussion.

\section{Numerical Integration of Stochastic Differential Equations}

A general system of continuous-time stochastic differential equations
(Langevin equations) can be written as 
\begin{equation}
\dot{x}_i = F_i(x_1,\cdots,x_n) + \sigma_{ij}(x_1,\cdots,x_n) \xi_j(t)
\label{geneq}
\end{equation}
where $i = 1, \cdots, n$ and $\xi_j(t)$ is a Gaussian white noise with 
\begin{eqnarray}
\langle\xi_j(t)\rangle & = & 0  \label{noise1}\\
\langle\xi_j(t) \xi_j(t')\rangle & = & \delta(t-t') \label{noise2}
\end{eqnarray}
and the symbol $\langle\cdots\rangle$ represents an average over
realizations of the inscribed variable (ensemble average). The noise
is said to be additive when $\sigma_{ij}$ is not a function of the
$x_i$, otherwise it is said to be multiplicative. In the case of
multiplicative noises, a mathematical subtlety arises in interpreting
stochastic integrals, the so-called Ito-Stratonovich ambiguity
\cite{cwg}. It should be stressed that this is a point of mathematics
and not of physics. Once it is clear how a particular Langevin
equation has been derived and what it is supposed to represent, it
should either be free of this ambiguity (as in the case of the example
we study later) or it should be clear that there must exist two
different stochastic equations, one written in the Ito form, the other
in Stratonovich, both representing the same physical process and hence
yielding identical answers for the variables of interest. (Another way
to state this is that there should be only one unique Fokker-Planck
equation.) It is important to note that the vast majority of numerical
update schemes for Langevin equations use the Ito form of the
equation.

The integral representation of the set of equations (\ref{geneq}) is
\begin{eqnarray}
x_i(t) &=& x_i(0) + \int_0^t ds F_i(x_1(s),\cdots,x_n(s)) \nonumber\\
&&+ \int_0^t ds \sigma_{ij}(x_1(s),\cdots,x_n(s)) \xi_j(s)
\label{int}
\end{eqnarray}
where $x_i(0)$ is a given sharp initial condition at $t=0$. The
infinitesimal update form of this equation may be derived by
replacing $t$ with an infinitesimal time step $h$:
\begin{eqnarray}
x_i(h) & = & x_i(0) + \int_0^h dt' \ F_i \left [x_k(0)+\int_0^{t'} ds 
F_k(x(s)) \right.\nonumber\\
&&\left. + \int_0^{t'} ds \sigma_{kl}(x(s)) \xi_l(s) \right ]
\nonumber\\ 
&& + \int_0^h dt' \ \sigma_{ij} \left [ x_k(0)+ \int_0^{t'} ds
F_k(x(s)) \right. \nonumber\\
&& \left.+ \int_0^{t'} ds \sigma_{kl}(x(s)) \xi_l(s) \right] \xi_j(t') 
\end{eqnarray}
Since $F_i$ and $\sigma_{ij}$ are smooth functions of the $x_i$, they
may be expanded about their values at $t=0$, in which case we can
write the exact solution for $x_i(h)$ as
\begin{equation}
x_i(h) = D_i(h) + S_i(h)
\end{equation} 
where $D_i(h)$ and $S_i(h)$ denote the deterministic and stochastic
contributions respectively. The deterministic contribution $D_i(h)$ is
\begin{equation}
D_i(h) = x_i(0) + h F_i + \frac{1}{2}h^2 F_{i,k} F_{k} + O(h^3)
\end{equation}
where $F_{i,k} = \partial F_i /\partial x_k$, the summation convention
for the repeated indices having being employed. The stochastic
contribution $S_i(h)$ is 
\begin{eqnarray}
&& S_i(h) = \sigma_{ij}W_j(h) + \sigma_{ij,k} \sigma_{kl} C_{lj}(h)
+F_{i,k}\sigma_{kl}Z_l(h) \nonumber\\
&& + \sigma_{ij,k}F_k(hW_j(h)-Z_j(h)) +
\frac{1}{2}\sigma_{ij,kl}\sigma_{km}\sigma_{ln}H_{mnj}(h) \nonumber\\
&& +\frac{1}{2}F_{i,kl}\sigma_{ks}\sigma_{lt}G_{st}(h) + \frac{1}{2} 
F_k\sigma_{ij,kl}\sigma_{lm}K_{mj}(h) \nonumber \\
&& + \frac{1}{2}F_l\sigma_{ij,kl}
\sigma_{km} K_{mj}(h) + \frac{1}{6} \sigma_{ij,klm}\sigma_{kn}
\sigma_{lo}\sigma_{mp}I_{nopj} \nonumber\\
&& + O(h^{5/2})
\label{stochcont}
\end{eqnarray}
The quantities $W_i$, $C_{ij}$, $H_{ijk}$, $Z_i$, $G_{ij}$, $K_{ij}$,
and $I_{ijkl}$ are random variables which can be written as stochastic
integrals over the Gaussian white noise $\xi(t)$:
\begin{eqnarray}
W_i(h) & = & \int_0^h dt \xi_i(t) \sim O(h^{1/2}) \label{stintbegin}\\
C_{ij}(h) & = & \int_0^h dt W_i(t) \xi_j(t) \sim O(h) \\ 
H_{ijk}(h) & = & \int_0^h dt W_i(t) W_j(t) \xi_k(t) \sim O(h^{3/2})\\ 
Z_i(h) & = & \int_0^h dt W_i(t) \sim O(h^{3/2}) \\ 
G_{ij}(h) & = & \int_0^h dt W_i(t) W_j(t) \sim O(h^2) \\ 
K_{ij}(h) & = & \int_0^h t dt W_i(t)\xi_j(t) \sim O(h^2) \\ 
I_{ijkl}(h) & = & \int_0^h dt W_i(t) W_j(t) W_k(t) \xi_l(t) \sim
O(h^2)\label{stintend} 
\end{eqnarray}
Ito integration has been employed in the derivation of the above equations.

The $n$th moment of the $x_i$ is 
\begin{eqnarray}
\langle x_i(h)^n\rangle & = & \langle(D_i(h) + S_i(h))^n\rangle  \nonumber \\ 
&  = & D_i(h)^n + nD_i(h)^{n-1}\langle S_i(h)\rangle\nonumber\\
&& +C_n^2 D_i(h)^{n-2}\langle(S_i(h))^2\rangle + \cdots
\label{mom}
\end{eqnarray}
where
\begin{eqnarray}
C_n^i & = & \left( \begin{array}{c}
             i \\
             n
         \end{array} \right )  = \frac{n!}{i! (n-i)!}
\end{eqnarray}
and 
\begin{eqnarray}
\langle S_i(h)\rangle & = & \frac{1}{4} F_{,kl}^i \sigma^{ks}
\sigma^{ls} h^2  + O(h^3)\label{mombegin}\\ 
\langle S_i(h)S_j(h)\rangle & = & \sigma^{il}\sigma^{jl} h +
\frac{1}{2} \sigma_{,k}^{im} \sigma^{kl} \sigma_{,p}^{jm} \sigma^{pl}
h^2 \nonumber\\
&& + \frac{1}{2} \sigma^{il} F_{,k}^j \sigma^{kl} h^2 +
\frac{1}{2}\sigma^{jl}F_{,k}^i \sigma^{kl} h^2  \nonumber\\
&& + \frac{1}{2}\sigma^{il}\sigma_{,k}^{jl}F^{k} h^2 +
\frac{1}{2}\sigma^{jl}\sigma_{,k}^{il} F^k h^2\nonumber \\ 
&& + \frac{1}{4}\sigma^{ip}\sigma_{,kl}^{jp}\sigma^{km} \sigma^{lm}
h^2\nonumber\\ 
&& + \frac{1}{4}\sigma^{jp}\sigma_{,kl}^{ip}\sigma^{km} \sigma^{lm}
h^2 + O(h^3)  \\
\langle S_i(h)S_j(h)S_k(h)\rangle & = & O(h^3) \\
\langle S_i(h)^4\rangle  & = & 3 (\sigma^{ii})^4 + O(h^3) \\
\langle (S_i(h))^5\rangle & = & O(h^3) \label{momend}
\end{eqnarray}
Suppose that the results from a numerical algorithm were written as
\begin{equation}
\bar{x}_i(h) = \bar{D}_i(h) + \bar{S}_i(h)
\end{equation} 
where the $\bar{x}_i$ are approximations to the exact solutions
$x_i$. The $n$th moment of $\bar{x}_i$ is
\begin{eqnarray}
\langle\bar{x}_i(h)^n\rangle & = & \langle (\bar{D}_i(h) +
\bar{S}_i(h))^n\rangle \nonumber\\
&=& \bar{D}_i(h)^n + n\bar{D}_i(h)^{n-1}\langle\bar{S}_i(h)\rangle
\nonumber \\  
&& +C_n^2\bar{D}_i(h)^{n-2}\langle (\bar{S}_i(h))^2\rangle + \cdots
\label{approx}
\end{eqnarray}
Comparing Eqns.~(\ref{mom}) and (\ref{approx}), we see that if
$D_i(h)$ and $\bar{D}_i(h)$, and $S_i(h)$ and $\bar{S}_i(h)$ coincide
up to $h^2$, we will have 
\begin{equation}
x_i(h) - \bar{x_i}(h) = O(h^3)
\end{equation}
and for a finite time interval
\begin{equation}
\langle x_i(t)^n\rangle - \langle \bar{x_i}(t))^n\rangle = O(h^2) 
\end{equation}

\section{Stochastic Leap-frog Algorithm For Brownian Motion}

The approach to modeling Brownian motion that we consider here is that
of a particle coupled to the environment through its position variable
\cite{zwanzig}. When this is the case, noise terms enter only in the
dynamical equations for the particle momenta. In the case of three
dimensions, the dynamical equations take the general form:
\begin{eqnarray}
\dot{x}_1 & = & F_1(x_1,x_2,x_3,x_4,x_5,x_6) +
\sigma_{11}(x_2,x_4,x_6) \xi_1(t) \nonumber \\ 
\dot{x}_2 & = & F_2(x_1) \nonumber \\
\dot{x}_3 & = & F_3(x_1,x_2,x_3,x_4,x_5,x_6) +
\sigma_{33}(x_2,x_4,x_6) \xi_3(t) \nonumber \\
\dot{x}_4 & = & F_4(x_3) \nonumber \\
\dot{x}_5 & = & F_5(x_1,x_2,x_3,x_4,x_5,x_6) +
\sigma_{55}(x_2,x_4,x_6) \xi_5(t) \nonumber \\
\dot{x}_6 & = & F_6(x_5) 
\label{bmeqs}
\end{eqnarray}
The convention used here is that the odd indices correspond to
momenta, and the even indices to the spatial coordinate. In the
dynamical equations for the momenta, the first term on the right hand
side is a systematic drift term which includes the effects due to
external forces and damping. The second term is stochastic in nature
and describes a noise force which, in general, is a function of
position. The noise $\xi(t)$ is first assumed to be Gaussian and white
as defined by Eqns.~(\ref{noise1})-(\ref{noise2}). The stochastic
leap-frog algorithm for the Eqns.~(\ref{bmeqs}) is written as
\begin{eqnarray}
\bar{x}_i(h) & = & \bar{D}_i(h) + \bar{S}_i(h)
\end{eqnarray}
The deterministic contribution $\bar{D}_i(h)$ can be obtained using
the deterministic leap-frog algorithm. The stochastic contribution
$\bar{S}_i(h)$ can be obtained by applying Eq.~(\ref{stochcont}) on
Eq.~(\ref{bmeqs}). The stochastic integration defined by
Eqs.~(\ref{stintbegin}) to (\ref{stintend}) can be approximated 
so that the moment relationships defined by Eqs.~(\ref{mombegin}) to
(\ref{momend}) are satisfied. After some calculation, the
deterministic contribution $\bar{D}_i(h)$ and the stochastic
contribution $\bar{S}_i(h)$ of the above recursion formula for
one-step integration are found to be 
\begin{eqnarray}
\bar{D}_i(h) & = & \bar{x}_i(0) + 
h F_i(\bar{x}_1^*,\bar{x}_2^*,\bar{x}_3^*,\bar{x}_4^*,\bar{x}_5^*,
\bar{x}_6^*);  \mbox{ \hspace{0.5cm}} \nonumber\\
&& \{i=1,3,5\} \nonumber \\
\bar{D}_i(h) & = & \bar{x}_i^* \nonumber\\
&& + \frac{1}{2} h
F_i\left[x_{i-1}+hF_{i-1}(\bar{x}_1^*,\bar{x}_2^*,\bar{x}_3^*,
\bar{x}_4^*,\bar{x}_5^*, 
\bar{x}_6^*)\right];\nonumber\\
&& \{i=2,4,6\} \nonumber \\
\bar{S}_i(h) & = &  \sigma_{ii}\sqrt{h} W_i(h) +
\frac{1}{2}F_{i,k}\sigma_{kk} h^{3/2} \tilde{W}_i(h)\nonumber\\
&& +\frac{1}{2} \sigma_{ii,j}F_j h^{3/2} \tilde{W}_i(h) \nonumber \\
& & +\frac{1}{4}F_{i,kl}\sigma_{kk}\sigma_{ll} h^2 \tilde{W}_i(h)
\tilde{W}_i(h); \nonumber\\
&& \{i=1,3,5;~ j=2,4,6;~ k,l = 1,3,5\}
\nonumber \\
\bar{S}_i(h) & = &  
\frac{1}{\sqrt{3}}F_{i,j}\sigma_{jj}h^{3/2}\tilde{W}_j(h) \nonumber\\
&&+\frac{1}{4}F_{i,jj}\sigma_{jj}^2
h^2\tilde{W}_j(h)\tilde{W}_j(h)\nonumber\\   
&&\{i=2,4,6;~ j=1,3,5\}  \nonumber \\
\bar{x}_i^* & = & \bar{x}_i(0) + \frac{1}{2} h F_i(\bar{x}_1,\bar{x}_2,
\bar{x}_3,\bar{x}_4,\bar{x}_5,\bar{x}_6)\nonumber\\
&& \{i = 1,2,3,4,5,6\} 
\label{walgo}
\end{eqnarray}
where $\tilde{W}_i(h)$ is a series of random numbers with the moments
\begin{eqnarray}
\langle\tilde{W}_i(h)\rangle &=&\langle(\tilde{W}_i(h))^3\rangle =
\langle(\tilde{W}_i(h))^5\rangle = 0  \\  
\langle(\tilde{W}_i(h))^2 \rangle & = & 1,~~~\langle(\tilde{W}_i(h))^4
\rangle = 3 
\end{eqnarray}
This can not only be achieved by choosing true Gaussian random
numbers, but also by using the sequence of random numbers following:
\begin{eqnarray}
\tilde{W}_i(h) & = & \left \{ \begin{array}{ccc}
                    -\sqrt{3 }, &  & R < 1/6 \\
                    0, &  & 1/6 \le R < 5/6  \\
                    \sqrt{3 }, &  & 5/6 \le R
                    \end{array} \right.
\end{eqnarray}
where $R$ is a uniformly distributed random number on the interval
(0,1). This trick significantly reduces the computational cost in
generating random numbers. 

Next we consider the case that the noise in Eqs.~(\ref{bmeqs}) is
a colored Ornstein-Uhlenbeck process which obeys
\begin{eqnarray}
\langle \xi_i(t)\rangle & = & 0 \\
\langle \xi_i(t)\xi_i(t') \rangle& = & \frac{k_i}{2} \exp(-k_i|t-t'|) 
\end{eqnarray}
where the correlation factor $k_i$ is the reciprocal of the
correlation time. In the limit of $k_i\rightarrow \infty$, the
Ornstein-Uhlenbeck process reduces to Gaussian white noise. 
The above process can be generated by using a white Gaussian noise
from a stochastic differential equation
\begin{equation}
\dot{\xi}_i(t) = -k_i \xi_i(t) + k_i \zeta_i(t)
\end{equation}
where $\zeta_i(t)$ is a standard Gaussian white noise. The initial
value $\xi_i(0)$ is chosen to be a Gaussian random number with
$\langle \xi_i(0)\rangle=0$ and $\langle\xi_i(0)^2\rangle=k_i/2$.

For the stochastic process with colored noise, the leap-frog algorithm 
for Eqns.~(\ref{bmeqs}) is of the same form as that for white noise
(Cf. Eqn.~(\ref{walgo})), but with 
\begin{eqnarray}
\bar{D}_i(h) & = & \bar{x}_i(0) + 
h F_i(\bar{x}_1^*,\bar{x}_2^*,\bar{x}_3^*,\bar{x}_4^*,\bar{x}_5^*,
\bar{x}_6^*) \nonumber\\
&& + h \sigma_{ii}(\bar{x}_2^*,\bar{x}_4^*,\bar{x}_6^*) \xi_i^*;
\nonumber\\
&&\{i=1,3,5\} \nonumber \\
\bar{D}_i(h) & = & \bar{x}_i^* \nonumber\\
&& + \frac{1}{2} h F_i\left[\bar{x}_{i-1} + h F_{i-1}
(\bar{x}_1^*,\bar{x_2}^*,\bar{x}_3^*,\bar{x_4}^*,\bar{x}_5^*,
\bar{x_6}^*) \right.\nonumber\\ 
&& \left. + h \sigma_{i-1i-1}(\bar{x}_2^*,\bar{x}_4^*,\bar{x}_6^*) 
\xi_{i-1}^* \right]; \nonumber \\ 
&& \{i=2,4,6\} \nonumber\\
\bar{D}_{\xi_i}(h) & = & \xi_i(0) \exp(-k_i h); \nonumber\\
&& \{i=1,3,5\} \nonumber \\
\bar{S}_i(h) & = &  
\frac{1}{\sqrt{3}}
\sigma_{ii}(\bar{x}_2,\bar{x}_4,\bar{x}_6) k_i h^{3/2} \tilde{W}_i(h);
\nonumber\\
&& \{i = 1,3,5\}  \nonumber \\
\bar{S}_i(h) & = &  0;  \nonumber\\
&&\{i=2,4,6\}\nonumber \\
\bar{S}_{\xi_i} & = & k_i \sqrt{h} \tilde{W}_i(h) - 
\frac{1}{2} k_i^2 h^{3/2} \tilde{W}_i(h);\nonumber\\
&&\{i = 1,3,5\}
\label{calgo}
\end{eqnarray}
where
\begin{eqnarray}
\bar{x}_i^* & = & \bar{x}_i(0) + \frac{1}{2} h \left[F_i(\bar{x}_1,\bar{x}_2,
\bar{x}_3,\bar{x}_4,\bar{x}_5,\bar{x}_6) \right.\nonumber\\
&&\left. +\sigma_{ii}(\bar{x}_2, \bar{x}_4,\bar{x}_6) \xi_i\right];
\nonumber\\
&&\{i=1,3,5\} \nonumber  \\
\bar{x}_i^* & = & \bar{x}_i(0) + \frac{1}{2} h F_i(\bar{x}_1,\bar{x}_2,
\bar{x}_3,\bar{x}_4,\bar{x}_5,\bar{x}_6); \nonumber\\
&&\{i=2,4,6\} \nonumber  \\
\xi_i^* & = & \xi_i(0) \exp(-\frac{1}{2} k_i h); \nonumber\\
&&\{i=1,3,5\}
\end{eqnarray}

\begin{figure}
\centerline{\epsfig{figure=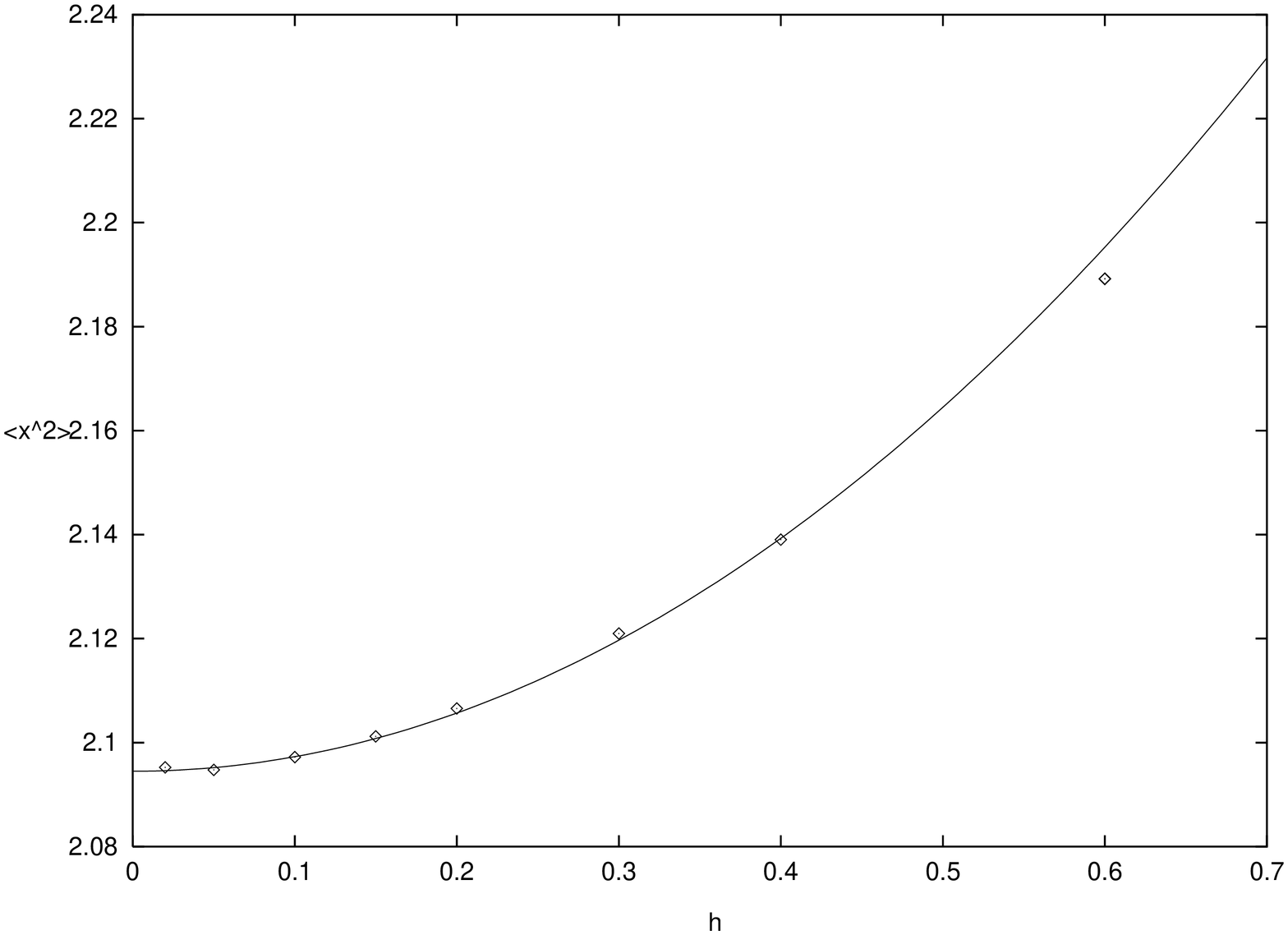,height=8cm,width=4cm,angle=-90}} 
\vspace{.5cm}
\centerline{\epsfig{figure=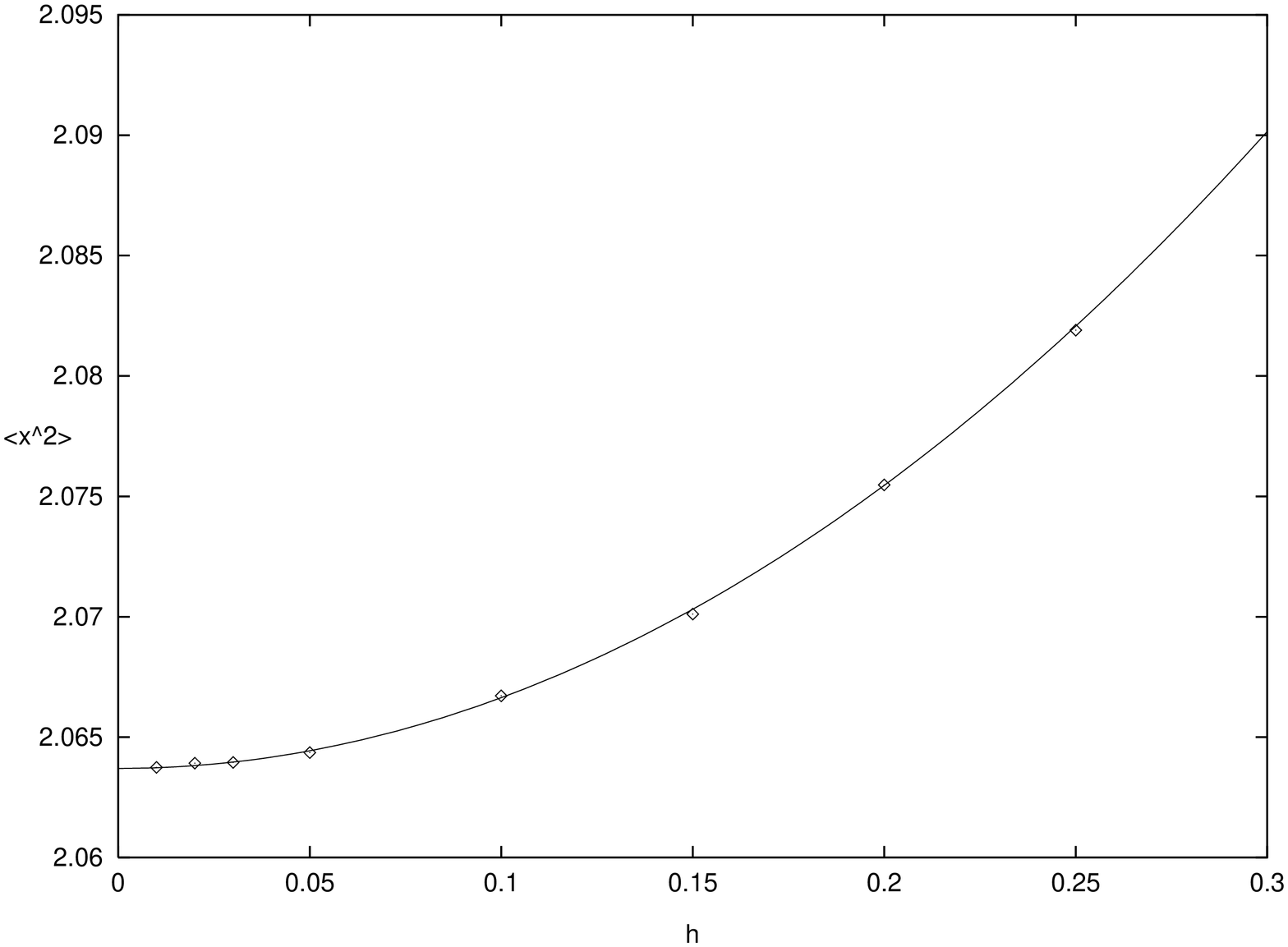,height=8cm,width=4cm,angle=-90}} 
\vspace{.5cm}
\caption{Zero damping convergence test. Top: $\langle x^2(t)\rangle$
at $t=6$ as a function of step size with white Gaussian noise. Bottom:
$\langle x^2(t)\rangle$ at $t=6$ as a function of step size with
colored Ornstein-Uhlenbeck noise. Solid lines represent quadratic fits
to the data points (diamonds).}
\label{quad}
\end{figure}

\section{Numerical Tests}
The above algorithms were tested on a one-dimensional stochastic
harmonic oscillator with a simple form of the multiplicative
noise. The equations of motion were 
\begin{eqnarray}
\dot{p} & = & F_1(p,x) + \sigma(x) \xi(t) \nonumber\\
\dot{x} & = & p
\label{testeqn}
\end{eqnarray}
where $F_1(p,x) = -\gamma p -\eta^2 x$ and $\sigma(x)=-\alpha x$.

As a first test, we computed $\langle x^2\rangle$ as a function of
time step size. To begin, we took the case of zero damping constant
($\gamma=0$), where $\langle x^2\rangle$ can be determined
analytically. The top curve in Fig.~\ref{quad} shows $\langle
x^2\rangle $ at $t=6.0$ as a function of time step size with white
Gaussian noise. Here, the parameters $\eta$ and $\alpha$ are set to
$1.0$ and $0.1$. The ensemble averages were taken over $10^6$
independent simulations.  The analytically determined value of
$\langle x^2\rangle $ at $t=6.0$ is $2.095222$ (The derivation of the
analytical results is given in the Appendix). The quadratic
convergence of the stochastic leap-frog algorithm is clearly seen in
the numerical results.  We then considered the case of colored
Ornstein-Uhlenbeck noise as a function of time step size using the
same parameters as in the white Gaussian noise case and with the
correlation parameter $k=0.16$. The result is shown as the bottom
curve in Fig.~\ref{quad} and the quadratic convergence is again
apparent.

\begin{figure}
\centerline{\epsfig{figure=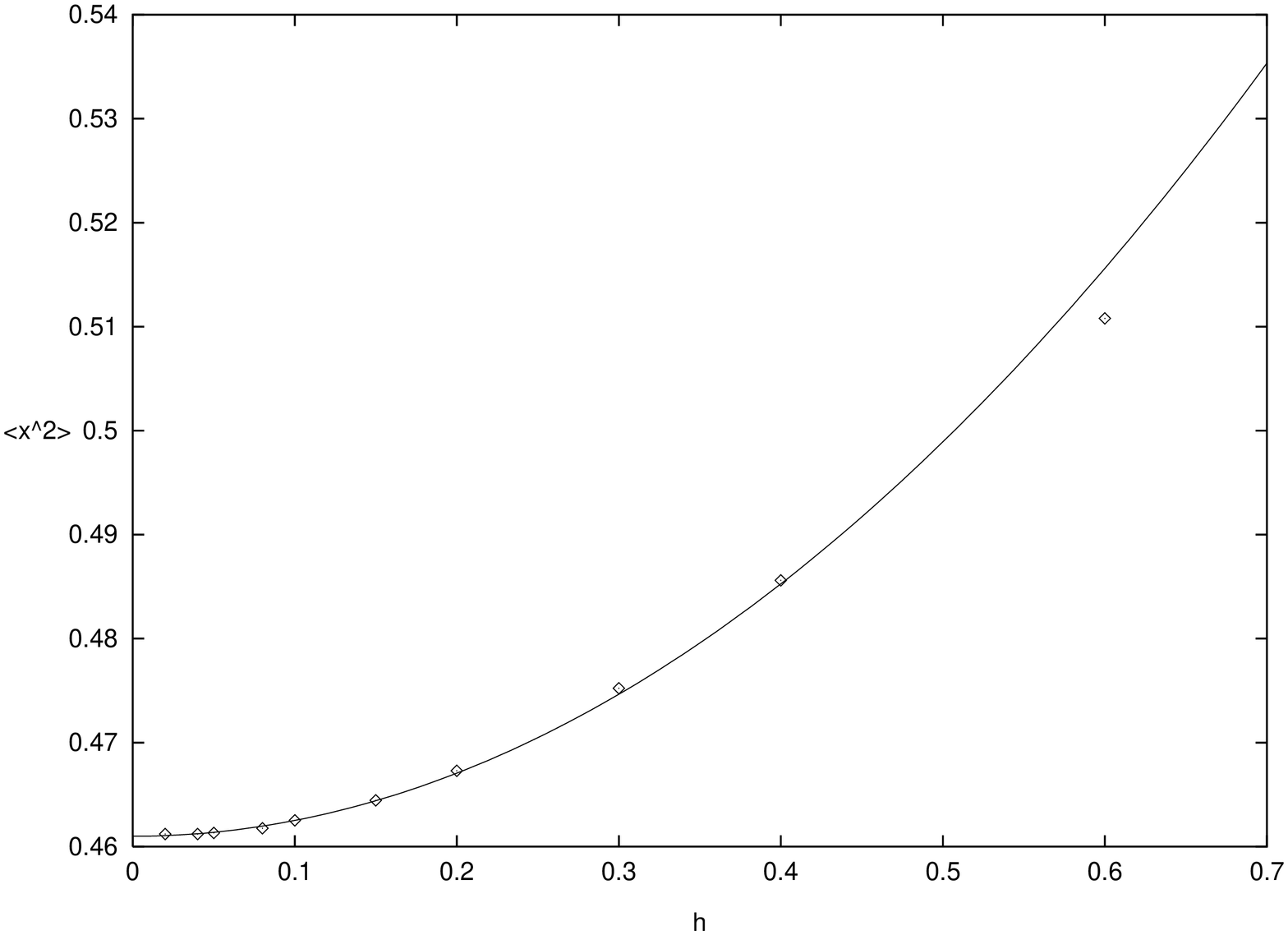,height=8cm,width=4cm,angle=-90}} 
\vspace{.5cm}
\centerline{\epsfig{figure=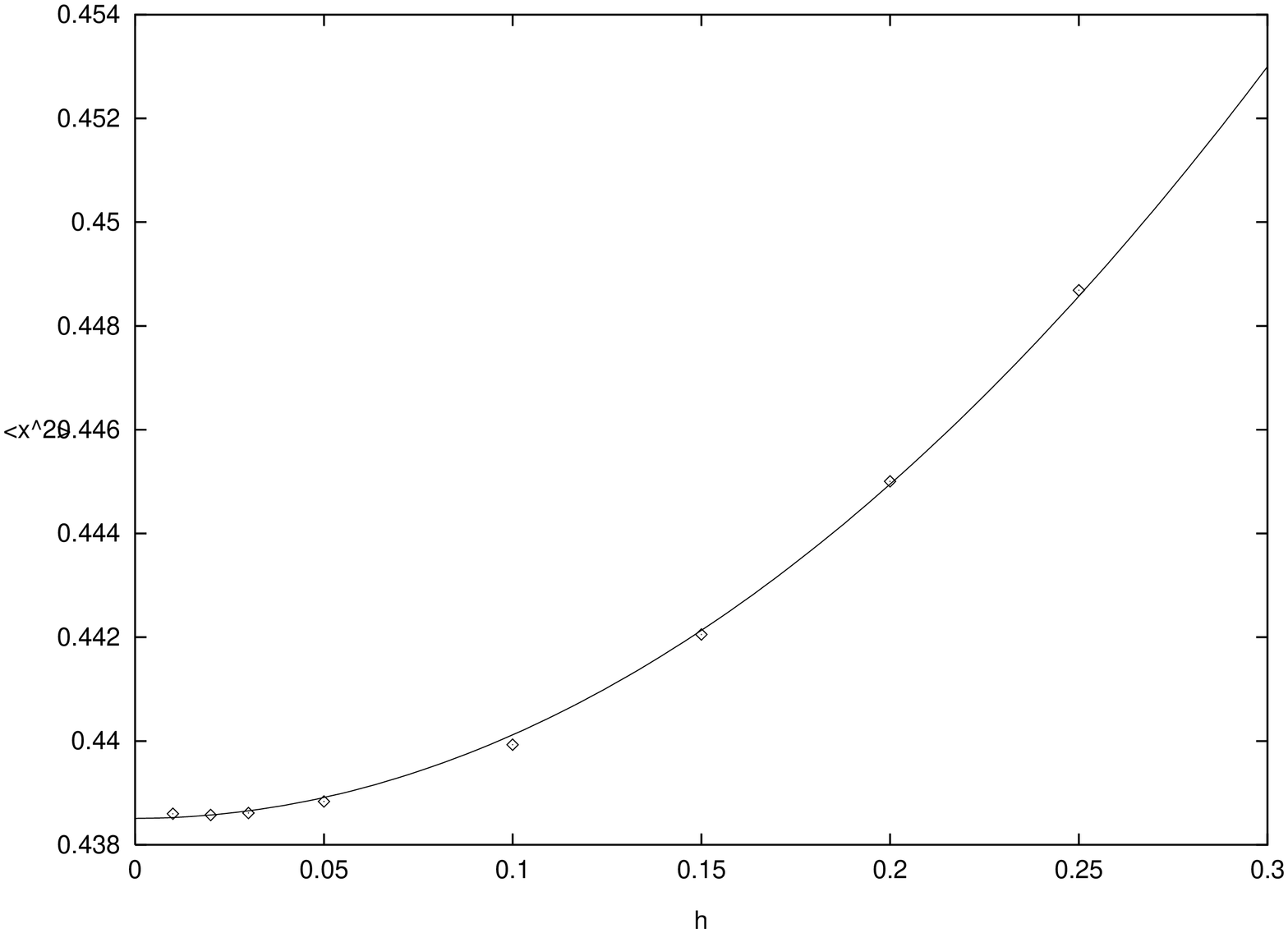,height=8cm,width=4cm,angle=-90}} 
\vspace{.5cm}
\caption{Finite damping ($\gamma=0.1$) convergence test. Top: $\langle
x^2(t)\rangle$ at $t=12$ as a function of step size with white
Gaussian noise. Bottom: $\langle x^2(t)\rangle$ at $t=12$ as a
function of step size with colored Ornstein-Uhlenbeck noise. Solid
lines represent quadratic fits to the data points (diamonds).}
\label{quad_damp}
\end{figure}

We verified that the quadratic convergence is present for nonzero
damping ($\gamma=0.1$). At $t=12.0$, and with all other parameters as
above, the convergence of $\langle x^2 \rangle$ as a function of time
step is shown by the top and bottom curves in Fig.~\ref{quad_damp}
(white Gaussian noise and colored Ornstein-Uhlenbeck noise,
respectively). 

\begin{figure}
\centerline{\epsfig{figure=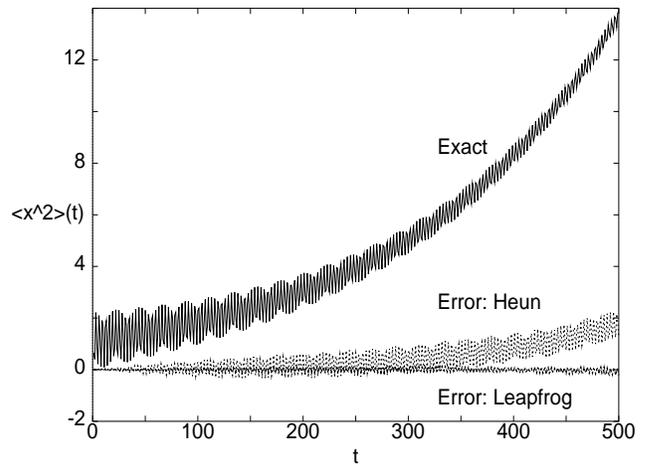,height=8cm,width=6cm,angle=-90}} 
\vspace{.5cm}
\caption{Comparing stochastic leap-frog and the Heun algorithm:
$\langle x^2(t)\rangle$ as a function of $t$. Errors are given
relative to the exact solution.}
\label{comp}
\end{figure}

As a comparison against the conventional Heun's algorithm, we computed
$\langle x^2 \rangle $ as a function of $t$ using $100,000$ numerical
realizations for a particle starting from $(0.0,1.5)$ in the $(x,p)$
phase space. The results along with the analytical solution and a
numerical solution using Heun's algorithm are given in
Fig.~\ref{comp}. Parameters used were $h=0.1$, $\eta=1.0$, and
$\alpha=0.1$. The advantage in accuracy of the stochastic leap-frog
algorithm over Heun's algorithm is clearly displayed, both in terms of
error amplitude and lack of a systematic drift.

We note that while in general Heun's algorithm is only linear for
multiplicative noise applications, for the particular problem at
hand it turns out to be quadratic. This is due to a coincidence: the
stochastic term of $x$ does not contain $W(h)$ but does posses a
higher order term $h W(h)$. However, this higher order term has a
larger coefficient compared with our stochastic leap-frog algorithm,
and this accounts for the larger errors observed in Fig.~3.

\section{A Physical Application: The Mechanical Oscillator}
In this section, we apply our algorithm to studying the approach to
thermal equilibrium of an oscillator coupled nonlinearly to a heat
bath modeled by a set of noninteracting harmonic oscillators
\cite{zwanzig}. The nonlinear coupling leads to the introduction of
multiplicative noise into the system dynamics. Lindenberg and Seshadri
have pointed out that, at weak coupling, multiplicative noise may
significantly enhance the equilibration rate relative to the rate for
weak linear coupling (additive noise) \cite{lindenberg}. We will
choose the same form of the coordinate couplings as in
Ref.~\cite{lindenberg}, in which case the additive noise equations are
\begin{eqnarray}
\dot{p} & = & -\omega_0^2 x - \lambda_0 p + \sqrt{2 D_0} \xi_0(t)
\nonumber\\ 
\dot{x} & = & p
\end{eqnarray}
and for the system with multiplicative noise only:
\begin{eqnarray}
\dot{p} & = & - \omega_0^2 x - \lambda_2 x^2 p
- \sqrt{2 D_0} x \xi_2(t) \nonumber\\ 
\dot{x} & = & p
\label{multle}
\end{eqnarray}
where the diffusion coefficients $D_i = \lambda_i k T, \ i = 0,2$,
$\lambda_i$ is the coupling constant, $k$ is Boltzmann's constant, $T$
is the heat bath temperature, and $\omega_0$ is the oscillator angular
frequency without damping. The approach to thermal equilibrium is
guaranteed for both sorts of noises by the fluctuation-dissipation
relation 
\begin{equation}
\langle \xi_i(t)\xi_j(s)\rangle=\delta_{ij}\delta(t-s)
\end{equation}
written here for the general case when both noises are simultaneously
present.  While in all cases, it is clear that the final distribution
is identical and has to be the thermal distribution, the precise
nature of the approach to equilibrium can certainly be different. We
wish to explore this issue in more detail. An important point to keep
in mind is that in this particular system of equations there is no
noise-induced drift in the Fokker-Planck equation obtained from the
Stratonovich form of the Langevin equation, i.e., there is no
Ito-Stratonovich ambiguity.

It is a simple matter to solve the Langevin equations given above
applying the algorithm from Eqs.~(\ref{walgo}). As our primary
diagnostic, we computed the noise-averaged energy $\langle
E(t)\rangle$ of the oscillator as a function of time $t$, where
\begin{equation}
E(t) = \frac{1}{2} p^2 + \frac{1}{2} \omega_0^2 x^2.
\end{equation} 
In the weak coupling limit and employing orbit-averaging (valid
presumably when the dynamical time scale is much smaller than the
relaxation time scale), one finds \cite{lindenberg}
\begin{equation}
\langle E(t)\rangle=kT-(kT-E_0)\hbox{e}^{-\lambda_0 t}
\end{equation}
in the case of additive noise (a result which can also be directly
obtained as a limiting case from the known form of the exact solution
given, e.g., in Ref. \cite{risken}). The corresponding form of
the approximate solution in the case of multiplicative noise is
\begin{equation}
\langle E(t)\rangle={E_0 kT\over E_0 + (kT-E_0)\exp(-\lambda_2
kTt/\omega_0^2)}.
\label{multe}
\end{equation}
While in the case of additive noise, the exponential nature of the
relaxation is already clear from the form of the exact solution
(cf. Ref. \cite{risken}), the situation in the case of multiplicative
noise is not obviously apparent as no exact solution is known to
exist. The prediction of a relaxation process controlled by a single
exponential as found in (\ref{multe}) is a consequence of the
assumption $\langle x^2(t) \rangle \simeq kT/\omega_0^2$ at ``late''
times, this implying a constant damping coefficient in the Langevin
equation (\ref{multle}).

The timescale separations necessary for the energy-envelope method to
be applicable are encoded in the following inequalities
\cite{lindenberg}:
\begin{eqnarray}
\frac{\lambda_0}{\omega_0} & \ll & 1; \mbox{\hspace{0.5cm} additive
noise }\\ 
\frac{kT \lambda_2}{\omega_0^3} & \ll & 1;
\mbox{\hspace{0.5cm} multiplicative noise} 
\label{multcond}
\end{eqnarray}
As a first check, we performed simulations with $\omega_0 = 1.0$,
$\lambda_0=\lambda_2=0.01$, and $kT=4.5$, in which case both the above
conditions are satisfied. Moreover, with these choices of parameter
values, and within the energy envelope approximation, the relaxation
time predicted for multiplicative noise is substantially smaller than
for the case of additive noise. At the same time we also ran a
simulation at $kT=200$ to see how the energy envelope approximation
for multiplicative noise breaks down at high temperatures.

\begin{figure}
\centerline{\epsfig{figure=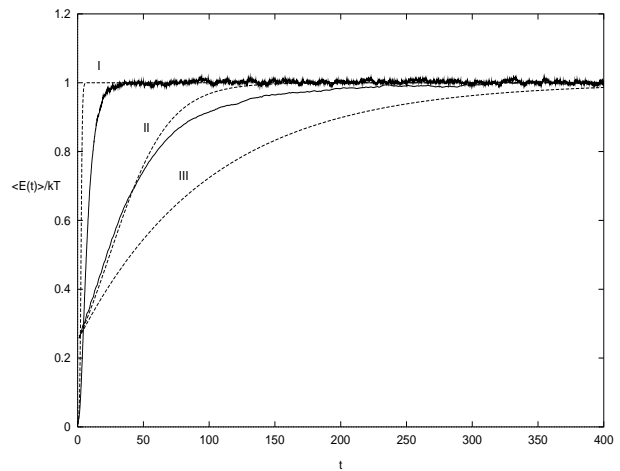,height=8cm,width=6cm,angle=-90}} 
\vspace{.5cm}
\caption{Temporal evolution of the scaled average energy $\langle
E(t)\rangle/kT$ with additive noise and multiplicative noise. The
dashed lines I and II are the predictions from Eqn. (\ref{multe}) for
$kT=200$ and $kT=4.5$ respectively. The dashed line III is the
theoretical prediction for additive noise with $kT=4.5$. As predicted,
the relaxation proceeds much faster with multiplicative noise: The solid
lines are numerical results for multiplicative noise at $kT=200$ and
$kT=4.5$. It is clear that at higher temperatures, the theory grossly
underestimates the relaxation time.} 
\label{app1}
\end{figure}

In Fig.~\ref{app1}, we display the time evolution of the average
energy (scaled by $kT$ for convenience) with additive and
multiplicative noise both from the simulations and the approximate
analytical calculations. In the case of weak coupling to the
environment (small $\lambda_0,~\lambda_2$), the rate at which the
average energy approaches equilibrium is significantly greater for the
case of multiplicative noise relative to the case of additive noise
more or less as expected. In addition, the analytic approximation
resulting from the application of the energy-envelope method
(\ref{multe}) is seen to be in reasonable agreement with the numerical
simulations for $kT=4.5$. The slightly higher equlibration rate from
the analytical calculation is due to the truncation in the energy
envelope equation using the $\langle E^2(t)\rangle\approx 2\langle
E(t)\rangle^2$ relation which yields an upper bound on the rate of
equilibration of the average energy~\cite{lindenberg}. Note that in
the case of high temperature ($kT=200$) the relaxaton time computed
from the energy envelope method is much smaller than the numerical
result, consistent with the violation of the condition
({\ref{multcond}).

While the results shown in Fig.~\ref{app1} do show that the energy
envelope approximation is qualitatively correct within its putative
domain of validity, it is clear that the actual relaxation process is
not of the precise form (\ref{multe}). In Fig.~\ref{app2} we
illustrate this point by plotting
\begin{equation}
{E_0(kT-\langle E(t)\rangle) \over \langle E(t)\rangle
(kT-E_0)}=\exp(-\lambda_2 kTt/\omega_0^2) 
\label{lhsrel}
\end{equation}
[equivalent to (\ref{multe})] against time on a log scale: the
relaxation is clearly nonexponential. The reason for the failure of
the approximation is that despite the fact that equipartition of
energy does take place on a relatively short time scale, it is not
true that $\langle x^2(t) \rangle$ can be treated as a constant even
at relatively late times.

\begin{figure}
\centerline{\epsfig{figure=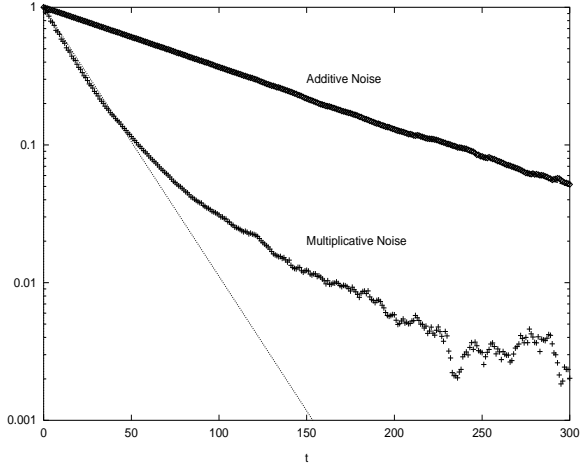,height=8cm,width=6cm,angle=-90}} 
\vspace{.5cm}
\caption{The LHS of (\ref{lhsrel}) as a function of time (straight
line) compared with numerical results for $kT=4.5$. Also shown is a
numerical result for the case of additive noise which is in excellent
agreement with the predicted exponential relaxation with the
relaxation timescale $=1/\lambda_0$.}
\label{app2}
\end{figure}

\section{Conclusions}
We have presented a stochastic leap-frog algorithm for single particle
Brownian motion with multiplicative noise. This method has the
advantages of retaining the symplectic property in the deterministic
limit, ease of implementation, and second-order convergence of moments
for multiplicative noise. Sampling a uniform distribution instead of a
Gaussian distribution helps to significantly reduce the computational
cost.  A comparison with the conventional Heun's algorithm highlights
the gain in acuracy due to the new method. Finally, we have applied
the stochastic leap-frog algorithm to a nonlinearly coupled
oscillator-heat-bath system in order to investigate the effect of
multiplicative noise on the nature of the relaxation process.

\section{Acknowledgments}
We acknowledge helpful discussions with Grant Lythe and Robert Ryne.
Partial support for this work came from the DOE Grand Challenge in
Computational Accelerator Physics. Numerical simulations were
performed on the SGI Origin2000 systems at the Advanced Computing
Laboratory (ACL) at Los Alamos National Laboratory, and on the Cray
T3E at the National Energy Research Scientific Computing Center
(NERSC) at Lawrence Berkeley National Laboratory.

\appendix
\section{}

The analytic solution of Eqns.~(\ref{testeqn}) for $\langle x^2(t)
\rangle$ (with white Gaussian noise) as a function of time in the
special case of zero damping, i.e. $\gamma = 0$, can be obtained by
solving the equivalent Fokker-Planck equation~\cite{risken} for the
probability density $f(x,p,t)$:
\begin{eqnarray} 
&&{\partial \over \partial t} f(x,p,t) = \nonumber\\
&& \left[ -p
{\partial \over \partial x} - {\partial F_1(p,x)\over \partial p}
+ {1\over 2} \sigma^2(x) {\partial^2 \over \partial p^2}\right]
f(x,p,t)   
\label{fpe}
\end{eqnarray} 
The expectation value of any function $M(x,p;t)$ can be written as
\begin{equation} 
\langle M(x,p)\rangle = \int_{- \infty}^{+\infty} d x d p M(x,p)
f(x,p,t) 
\label{momgen}
\end{equation}
Equations (\ref{fpe}) and (\ref{momgen}) can be used to yield a
BBGKY-like heirarchy for the evolution of phase space moments. Since
the system we are considering is linear, this heirarchy truncates
exactly and yields a group of coupled linear ordinary differential
equations for the moments $\langle x^2\rangle$, $\langle x
p\rangle$, and $\langle p^2\rangle$. These equations can be
written as a single third-order time evolution equation for $\langle
x^2\rangle$:  
\begin{equation} 
\frac{d^3\langle x^2\rangle}{d t^3} = -4 \eta^2 \frac{d \langle
x^2\rangle}{d t} + 2 \alpha^2 \langle x^2\rangle
\end{equation} 
subject to the initial conditions 
\begin{eqnarray} 
\langle x^2(0)\rangle & = & x^2(0)  \nonumber \\ 
{\langle \dot{x^2}(0)\rangle} & = & 2 x(0) p(0) \nonumber\\ 
{\langle \ddot{x^2}(0)\rangle} & = & 2p^2(0) - 2 \eta^2 x^2(0) 
\end{eqnarray} 
This equation has an analytical solution written as
\begin{equation} 
\langle x^2(t)\rangle= c_1 \exp(r_1 t) + c_2 \exp(r_2 t) + c_3
\exp(r_3 t)  
\end{equation} 
where $c_1$, $c_2$, and $c_3$ are constants depending on initial
conditions, and $r_1$, $r_2$ and $r_3$ are the roots of a third order
alegbraic equation 
\begin{equation} 
2 \alpha^2 - 4 \eta^2 x - x^3 = 0 
\end{equation} 
which gives 
\begin{eqnarray} r_1 & = & \left(\sqrt{64/27 \eta^6 + \alpha^4} +
\alpha^2 \right )^{1/3} \nonumber\\
&& - \left( \sqrt{64/27 \eta^6 + \alpha^4} - \alpha^2 \right )^{1/3}
\nonumber\\  
r_2 & = & \frac{1}{2}(1+\sqrt{3} i)\left( \sqrt{64/27 \eta^6 +
\alpha^4} - \alpha^2 \right )^{1/3} \nonumber\\
&& - \frac{1}{2}(1-\sqrt{3} i) \left(\sqrt{64/27 \eta^6 + \alpha^4} +
\alpha^2 \right )^{1/3}  \nonumber \\ 
r_3 & = & r_2^*
\end{eqnarray}
where the superscript $*$ represents complex conjugation. The positive
real root $r_1$ implies that $\langle x^2(t)\rangle $ will have an
exponential growth in time.

\end{document}